\documentclass[twocolumn,superscriptaddress,showpacs,aps,prb]{revtex4}
\usepackage{makeidx}
\usepackage{bm}
\usepackage{graphics}
\usepackage[dvips]{epsfig}
\newcounter{fig}

\begin{document}
\title{Symmetry constraints on phonon dispersion in graphene  }
\author{L.A. Falkovsky}
\affiliation{L.D. Landau Institute for Theoretical Physics, Moscow
117334, Russia} \affiliation{Institute of the High Pressure
Physics, Troitsk 142190, Russia}
\pacs{63.20.Dj, 81.05Uw, 71.15.Mb}
\begin{abstract}
Taking into account the constraints imposed by the lattice
symmetry, the phonon dispersion   is calculated for graphene with
interactions between the first and second nearest neighbors in the
framework of the Born-von Karman model. Analytical expressions are
obtained for the out-of-plane (bending) modes determined only by
two force constants as well as for the in-plane modes with four
force constants. Values of the force constants are found in
fitting to elastic constants and Raman frequencies observed in
graphite.
\end{abstract}
\maketitle
\section{Introduction}
Since the discovery of  graphene (a single atomic layer of
graphite)  \cite{Novo, ZSA}, main attention has been devoted to
its electronic properties. More recently, Raman spectroscopy
\cite{FMS} extends to investigations of graphene. For
interpretations of the Raman scattering as well as of the
transport phenomena, the  detailed knowledge of the lattice
dynamics and the electron-phonon interactions  is needed
\cite{NG}.

Several models \cite{DL,NWS,AR,NB,MKH,AD,GMR,LDC} have been
proposed to calculate the phonon dispersion in bulk graphite. Most
improved ones \cite{MKH,AD}  involve   many (up to twenty)
parameters.  Recently, detailed measurements and first-principles
calculations of optical phonon frequencies were made for graphite
\cite{MRTR,MMD}. They show the qualitative disagreement with the
models \cite{DL,LDC}, which include the central and angular atomic
forces between the first and second neighbors in the graphite
lattice.

The passage in the lattice dynamics from graphite to graphene and
then to nanotubes was examined in the {\it ab initio} calculation
\cite{SAS,DK}, in Ref. \cite{CMC} using the model \cite{DL}, and
in Ref. \cite{GSK} up to the fourth neighbor with 12 force
constants.  Numerical calculations based on H\"{u}ckel's theory
 \cite{MCZ}  and   in terms of the electron energy
were performed in work \cite{PL}. The first-principles
calculations \cite{MM} of dynamical properties for graphite and
graphene  (and also for diamond) show that distinctions between
the phonon frequencies in graphene and relating ones in graphite
are negligible in comparison with the experimental errors for that
frequencies in graphite. For the highest frequencies, this could
be intuitively expected because interactions between the adjacent
layers in graphite are weak.

Our aim here is to find an  analytical description   of the phonon
dispersion in graphene.  This can be done
 within the framework of  the  Born-von-Karman model for
 the honeycomb graphene lattice with interactions only between
  first and second nearest neighbors, but the
 constraints imposed by the lattice symmetry  should be taken
  into account.
 We show that the out-of-plane (bending)  and in-plane modes are
  decoupled from each other. The out-of-plane modes are described by two
  force constants  determined in fitting to
  the  Raman frequency and  smallest
  elastic constant $C_{44}$.
  In the narrow  wave-vector interval near the $\Gamma$ point,
  the acoustic out-of-plane mode has a linear dispersion with the
  non-zero sound velocity in  contrast to the so-called membrane mode
  which demonstrates the quadratic dispersion
   in the continuous-medium approximation \cite{LL}.
  We do not pay close attention to
  the agreement of lower frequencies with
  experiments on graphite because their values in graphene
   are convincingly less that ones in graphite. The four force
   constant describing the in-plane modes are found in fitting to
   their  Raman frequencies and elastic constant
   $C_{11}$ and $C_{12}$ of graphite.
The extent of agreement of the present theory with experiments
corresponds to the comparison level between the first-principal
calculations \cite{MRTR} and their data (see Table 3).
\section{ Phonon dynamics in nearest neighbor approximation }

  The equations of moution in the harmonic
approximation are written in the well-known form
\begin{equation} \label{eqmo}
\sum\limits_{j, m, \kappa'}\Phi^{\kappa \kappa'}_{ij}({\bf a}_n -
{\bf a}_m) u_j^{\kappa'} ({\bf a}_m)-\omega^2 u_i^{\kappa}({\bf
a}_n) = 0,
\end{equation}
where the vectors ${\bf a}_{n}$ numerate the lattice cells, the
superscripts $\kappa, \kappa'$  note two sublattices $A$ and $B$,
and the subscripts $i, j = x, y, z$ take three values
corresponding to the space coordinates. Since the potential energy
is the quadratic function of the atomic displacements
$u^{A}_{i}({\bf a}_n) $ and $u^{B}_{i}({\bf a}_n) $, the
force-constant matrix can be taken in the symmetric form,
$\Phi_{ij}^{A B}({\bf a}_{n}) = \Phi_{ji}^{B A}(-{\bf a}_{n}),$
and its Fourier transform, i.e. the dynamical matrix, is a
Hermitian matrix.

Each atom, for instance, ${\bf A}_0$ (see Fig. 1) has  three first
neighbors in the other sublattice, {\it i. e.}, $B$ with the
relative vectors
\[{\bf B}_{1}=a(1, 0),\quad {\bf B}_{2,3}=a(-1,\pm\sqrt{3})/2,\]
 and six second neighbors in the same sublattice $A$ with the relative vectors
\begin{eqnarray}\nonumber
{\bf A}_{1,4}=&\pm a(0, \sqrt3),\quad {\bf A}_{2,5}=\pm
a(-3,\sqrt3)/2,\\ \nonumber &{\bf A}_{3,6}=\mp a(3,\sqrt3)/2,
\end{eqnarray}
 where $a = 1.42 \AA$ is  the carbon-carbon distance.

For the nearest neighbors (in the $B$ sublattice),   the dynamical
matrix  has the form
\begin{eqnarray} \label{sc}
&\phi^{AB}_{i j}({\bf q}) = \sum_{\kappa=1}^3 \Phi^{AB}_{i j}({\bf
B}_{\kappa})\exp(i{\bf qB}_{\kappa}),
\end{eqnarray}
and
 for the next neighbors (in the $A$ sublattice)
\begin{eqnarray} \label{ss}
&\phi^{AA}_{i j}({\bf q})= \Phi^{AA}_{i j}({\bf
A}_{0})+\sum_{\kappa=1}^6 \Phi^{AA}_{i j}({\bf
A}_{\kappa})\exp(i{\bf qA}_{\kappa}),
\end{eqnarray}
where ${\bf A}_{0}$ indices the atom chosen at the center of the
coordinate system in the $A$ sublattice and the wave vector ${\bf
q}$ is taken in units of $1/a$.
\begin{figure}[h]\resizebox{.2\textwidth}{!}{\includegraphics{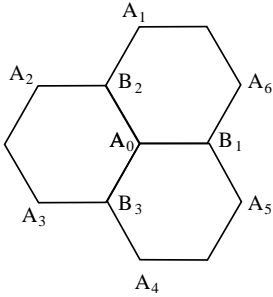}}
\caption{\label{1} First and second neighbors in the graphene
lattice.}
\end{figure}

The point  group $D_{6h}$ of the honeycomb lattice is generated by
$\{C_6,\sigma _v,\sigma _z\}$, where   $\sigma_ z$ is a reflection
$z\to -z$ by the plane that contains the graphene layer, $C_6$ is
a rotation by $\pi/3$ around the $z$ axis, and $\sigma_v$ is a
reflection by the $xz$ plane. The transformations of the group
impose constraints on the dynamical matrix. To obtain them, we
introduce variables $\xi,\eta=x\pm iy$ transforming under the
rotation $C_3$ around the $z$-axis
 (taken at the ${\bf A}_{0}$ atom) as follows
 $(\xi,\eta)\rightarrow (\xi,\eta)\exp(\pm 2\pi i/3)$. In
the  rotation, the atoms change their positions ${\bf
B}_1\rightarrow {\bf B}_2\rightarrow{\bf B}_3$,  ${\bf
A}_1\rightarrow {\bf A}_3\rightarrow{\bf A}_5$, and ${\bf
A}_2\rightarrow {\bf A}_4\rightarrow{\bf A}_6$. Therefore, all the
force constants $\Phi^{AB}_{\xi \eta}({\bf B}_{\kappa})$ with the
different $\kappa$ (as well as $\Phi^{AB}_{z z}({\bf
B}_{\kappa})$) are equal  to one another, but the force constants
with the coincident  subscripts $\xi$ or $\eta$  transform as
covariant variables. For instance,
\begin{eqnarray}\label{rc}
\Phi^{AB}_{\xi\xi}({\bf B}_{1})=\Phi^{AB}_{\xi\xi}({\bf
B}_{2})\exp{(2\pi i/3)}\\ \nonumber=\Phi^{AB}_{\xi\xi}({\bf
B}_{3})\exp{(-2\pi i/3)}. \end{eqnarray}

 The relation between
$\Phi^{AA}_{\xi\xi}({\bf A}_{\kappa})$ with the points ${\bf
A}_{1},{\bf A}_{3},{\bf A}_{5}$ (and also between  ${\bf
A}_{4},{\bf A}_{2},{\bf A}_{6}$)  has the same form. The constants
$\alpha_z=\Phi^{AB}_{zz}({\bf B}_{1}),
\gamma_z=\Phi^{AA}_{zz}({\bf A}_{1}),
\alpha=\Phi^{AB}_{\xi\eta}({\bf B}_{1}),$ and
$\gamma=\Phi^{AA}_{\xi\eta}({\bf A}_{1})$
are evidently real. The constant $\beta=\Phi^{AB}_{\xi\xi}({\bf
B}_{1})$ is real because the reflection $(x, y) \rightarrow (x,
-y)$ with  ${\bf
B}_{1}\rightarrow{\bf
B}_{1}$ belongs to the symmetry group. Besides,
for the first and second neighbors, we have one complex force
constant $\delta=\Phi^{AA}_{\xi\xi}({\bf A}_{1})$.

Two  force constants $\Phi^{AA}_{zz}({\bf A}_{0})$
 and $\Phi^{AA}_{\xi\eta}({\bf
A}_{0})$ for the atom ${\bf A}_{0} $  can be excluded with the
help of conditions imposed by invariance  with respect to the
translations of the layer as a whole in the $x/z$  directions.
Using the equations of motion (\ref{eqmo}) and  Eqs. (\ref{sc}),
(\ref{ss}), we find this stability condition
\begin{eqnarray}\label{ti}
  \Phi^{AA}_{\xi
\eta}({\bf A}_{0})+6\Phi^{AA}_{\xi\eta}({\bf
A}_{1})+3\Phi^{AB}_{\xi \eta}({\bf B}_{1})=0
\end{eqnarray}
and  the similar form for the $zz$ components.

\subsection {Dispersion of the bending out-of plane modes.}

In the first- and second-neighbor approximation, the out-of-plane
vibrations $u_z^{A}, u_z^{B}$ in the $z$ direction
 are not coupled with the in-plane modes because the force constants of  type
 $\Phi_{xz}$ or $\Phi_{yz}$ equals zero due to the reflection  $z\to -z$.
 The corresponding dynamical matrix
 for the out-of-plane modes has the form
\begin{equation}
\label{hamz} \left ( \begin{array}{cc}
 \phi^{AA}_{zz}({\bf q})&\phi^{AB}_{zz}({\bf q})\\
\phi^{AB}_{zz}({\bf q})^*&\phi^{AA}_{zz}({\bf q})
\end{array}\right ),
\end{equation}
where
\begin{eqnarray} \nonumber
&\phi^{AA}_{zz}({\bf q})=\\ \nonumber
&2\gamma_z[\cos{(\sqrt3q_y)}+2\cos{(3q_x/2)}
\cos{(\sqrt3q_y/2)}-3]-3\alpha_z,\\
\label{ag} &\phi^{AB}_{zz}({\bf q})=\\ \nonumber &\alpha_z
[\exp{(iq_x)}+2\exp{(-iq_x/2)} \cos{(\sqrt3q_y/2)}].
\end{eqnarray}
The phonon dispersion for the out-of-plane modes is found
\begin{equation} \label{pdz}
\omega_{\text{ZO,ZA}}({\bf q})=\sqrt{\phi^{AA}_{zz}({\bf q})\pm
|\phi^{AB}_{zz}({\bf q})|}.
\end{equation}
The equations allow  us to express the phonon frequencies of the
out-of plane branches at the critical points $\Gamma , K$, and $M$
in terms of the force constants:
\begin{eqnarray} \label{gz}
 &\omega_{\text{ZO}}(\Gamma)=\sqrt{-6\alpha_z} ,\quad
\omega_{\text{ZO, ZA}}(K)=\sqrt{-3\alpha_z-9\gamma_z},\\ \nonumber
&\omega_{\text{ZO, ZA}}(M)=\sqrt{(-3\mp 1)\alpha_z-8\gamma_z}.
\end{eqnarray}

 Expanding Eq. (\ref{pdz}) in powers of the wave vector $\bf{q}$, we find
  the velocity of the acoustic out-of-plane mode propagating in the
layer
\begin{equation}\label{sz}
s_{z}=a\left[-0.75\alpha_z- 4.5\gamma_z\right]^{1/2}=
\sqrt{C_{44}/\rho},
\end{equation}
where we use the well-known formula for the velocity of the
acoustic $z$-mode propagating in the $x$-direction in terms of the
elastic constant $C_{44}$ and  density $\rho$ of a hexagonal
crystal. Because the interaction between the layers in graphite is
weak, we can correspond the values of $C_{44}$ and   $\rho$  to
graphite. Using the values of $\alpha_z$ and $\gamma_z$ obtained
in fitting  to the exprimental data (see Table 1), we find the
value of the sound velocity for the out-plane mode $s_z=1.57\times
10^5$ cm/s.

\subsection {Dispersion of the in-plane modes.}

The dynamical matrix for the in-plane vibrations has the form
similar to that for the in-plane mode (\ref{hamz}), but instead of
the functions $\phi_{zz}^{AA}({\bf q})$ and $\phi_{zz}^{AB}({\bf
q})$ we have to substitute respectively the $2\times2$ matrices
\begin{equation}\label{hamm}
\phi^{AA}({\bf q})=\left ( \begin{array}{cc}
   \phi^{AA}_{\xi\eta}({\bf q})&\phi^{AA}_{\xi\xi}({\bf q})\\
   \phi^{AA}_{\xi\xi}({\bf q})^*&\phi^{AA}_{\xi\eta}({\bf q})
\end{array}\right ),
\end{equation}
\begin{equation}\label{hamb}
\phi^{AB}({\bf q})=\left ( \begin{array}{cc}
   \phi^{AB}_{\xi\eta}({\bf q})&\phi^{AB}_{\xi\xi}({\bf q})\\
   \phi^{AB}_{\eta\eta}({\bf q})&\phi^{AB}_{\xi\eta}({\bf q})
\end{array}\right ).
\end{equation}
The  matrix elements $\phi^{AA}_{\xi\eta}({\bf q})$ and
$\phi^{AB}_{\xi\eta}({\bf q})$ are obtained from $\phi^{AA}_{zz}({\bf q})$
and $\phi^{AB}_{zz}({\bf q})$, Eqs. (\ref{ag}), respectively, with
substitutions  $\gamma$ and $\alpha$ instead of $\gamma_z$ and
$\alpha_z$. The off-diagonal elements are given by
\begin{eqnarray}\label{od}
&\phi^{AA}_{\xi\xi}({\bf q})=\\\nonumber &\delta
[\exp(i\sqrt{3}q_y)+2\cos{(3q_x/2+2\pi/3)}\exp(-i\sqrt3q_y/2)]+\\
 \nonumber &\delta^*
[\exp(-i\sqrt{3}q_y)+2\cos{(3q_x/2-2\pi/3)}\exp{(i\sqrt3q_y/2)}],\\
\nonumber &\phi^{AB}_{\xi\xi}({\bf q})= \\\nonumber &\beta
[\exp{(iq_x)}+2\exp{(-iq_x/2)}
\cos{(\sqrt3q_y/2-2\pi/3)}].
\end{eqnarray}
The matrix elements for the $B$ sublattice  can be obtained from
 that ones for the $A$ sublattice  by  $C_2$ rotation
$(x,y)\rightarrow -(x,y)$ of the graphene symmetry group.

The optical phonon frequencies for the in-plane branches at
$\Gamma$ and $K$  are found
\begin{eqnarray}\label{pfx}\nonumber
\omega_{1,2}^{in-pl}(\Gamma)=\sqrt{-6\alpha},\quad
\text{doublet},\\
\omega^{in-pl}_{1,2}(K)=\sqrt{-3\alpha-9\gamma},\quad
\text{doublet},\\ \nonumber
\omega^{in-pl}_{3,4}(K)=\sqrt{-3\alpha-9\gamma\pm 3\beta}.
\end{eqnarray}
Using Eqs. (\ref{hamm})-(\ref{od}), we  find the   in-plane mode
dispersion in the explicit form for the $G-K$ direction. An
algebraic equation of the forth order have to be solved for the M
point as well as for points of the general position.

{

The in-plane vibrations make a contribution into the elastic
constants C$_{11}$ and C$_{12}$. The corresponding relation
between the dynamic matrix elements and the elastic constants  can
be deduced taking the long-wavelength limit (${\bf q}\rightarrow
0$) in the matrices (\ref{hamm}) and (\ref{hamb}). In this limit,
separating the acoustic vibrations ${\bf u}^{\text{ac}}$ from the
optical modes, we obtain the equation of motion in the matrix form
\begin{equation}\label{ac}\begin{array}{c}
\left[(\phi ^{AA}+\phi ^{AB}+ \phi ^{BB}+ \phi ^{BA})/2\right.\\+
\left.\phi _1^{AB} (\phi _0^{AB})^{-1} \phi _1^{AB}
-\omega^2\right] {\bf u}^{\text{ac}}=0,
\end{array}\end{equation}
where the subscripts 0 and 1 mean that  the terms of the zero and
first order in ${\bf q}$ should, correspondingly,  be kept in the
matrices (\ref{hamm}) and (\ref{hamb}) but the expansion to the
second order  is used in other terms. We  find the matrix factor
of ${\bf u}^{\text{ac}}$ in Eqs. (\ref{ac}):
\begin{equation}\label{acm}
 \left ( \begin{array}{cc} s_1q^2-\omega^2& s_2q_{+}^2\\
                                       s_2q_{-}^2
                                       &s_1q^2-\omega^2
\end{array}\right ), \end{equation}
where
\begin{equation}\label{s12}
s_1=-\frac{9}{2}\gamma-\frac{3}{4}(\alpha-\beta^2/\alpha),\quad
 s_2=\frac{9}{8}\delta-\frac{3}{8}\beta.
 \end{equation}
With the help of Eq. (\ref{acm}),  we obtain the  velocities of
longitudinal and transverse acoustic in-plane modes
\begin{equation}\label{lv}
 \begin{array}{c}
 s_{\text {LA}}=a\sqrt{s_1+s_2}=\sqrt{C_{11}/\rho},\,\\
 s_{\text {TA}}=a\sqrt{s_1-s_2}=\sqrt{(C_{11}-C_{12})/2\rho},
\end{array}\end{equation}
corresponding them to   elastic
 constants and  density of graphite. The values of  force
 constants from  Table 1 give the sound velocities
$s_{\text{ LA}}= 2.03\times 10^6 \text{cm/s},\, s_{\text {TA}}=
1.31\times 10^6 \text{cm/s}\,.$
\begin{table}[h]
\caption{\label{tb1} Force constants in 10$^{5}$  cm$^{-2}$ . }
        \begin{ruledtabular}
                \begin{tabular}{|c|c|c|c|c|c|}
     $\alpha$ &$\beta$&$\gamma$ &$\delta$   &$\alpha_z$ &$\gamma_z$\\
\hline
 -3.980& -1.132&-0.297&1.123&-1.270&0.204
\end{tabular}
\end{ruledtabular}
\end{table}
\begin{table}[h]
\caption{\label{tb1} Elastic constants (in 10 GPa) calculated
(theo) and observed \cite{GP} (exp). }
        \begin{ruledtabular}
                \begin{tabular}{|c|c|c|c|}
      &$C_{11}$ &$C_{12}$ &$C_{44}$\\
\hline
 theo&$ 93$ &$16$ &$0.56$\\
\hline
exp &$106\pm2$ &$18\pm2$ &$0.45\pm.05 $\\
\end{tabular}
\end{ruledtabular}
\end{table}
\begin{table*}[]
\caption{\label{tb1} Phonon frequencies at critical points in
cm$^{-1}$; $z$ and $\parallel$ stand for the out-of-plane and
in-plane branches, respectively. }
        \begin{ruledtabular}
                \begin{tabular}{|c|c|c|c|}
   &$\Gamma$ \, [0 0]    &$M$  \,   $[1 \sqrt3]\pi/3a$    &$K$\, $[0 1]4\pi/3\sqrt3 a$ \\
 \hline
&$\omega^{\parallel}$\qquad$\omega^{z}$&$\omega^{\parallel}_1$\qquad
$\omega^{\parallel}_2$\qquad$\omega^{\parallel}_3$\qquad$\omega^{\parallel}_4$\qquad$\omega^{z}_1$
\quad
$\omega^{z}_2$&$\omega^{\parallel}_1$\qquad $\omega^{\parallel}_{2,3}$\qquad $\omega^{\parallel}_4$\quad$\omega^{z}_{1,2}$\\
\hline exp&1590$^a$\quad 861$^a$\quad&\,1389$^a$\qquad\qquad\quad
\quad\quad
630$^d$\quad 670$^a$\quad 471$^c$ &1313$^d$\, 1184$^b$\quad \quad\quad 482$^d$\quad \\
      &1583$^b$\quad 868$^c$\quad&1390$^b $\, \,1323$^b \,$\,
1290$^b\,\,$\,\,
625$^e$\quad 625$^e$\quad 451$^d$ &1265$^b$\, 1194$^b$\quad \quad\quad 517$^d$\quad \\
      &1565$^b$\quad 874$^e$\quad&\qquad\qquad \qquad \qquad\quad
\quad\quad \qquad\qquad \,\,480$^e$ & 1285$^e$ \qquad\quad 1011$^e$\, 531$^e$ \\
\hline
             theo$^b$&\, 1581 \quad\qquad\qquad&1425\quad 1350\quad
\,1315\quad\,
\qquad\quad\quad\qquad\qquad\qquad& 1300\quad 1220\quad 950\qquad\quad\quad \\
\hline theo&1545 \quad873\quad&1432\quad 1307\quad \,1268\quad\,
\,598\,\quad587\qquad 301&1342\quad 1209\quad 1059\quad 444\quad \\
\end{tabular}
\end{ruledtabular}
 $^a$ Reference \cite{OA}, $^b$ Reference \cite{MRTR}, $^c$
Reference \cite{NWS}, $^d$ Reference \cite{Y}, $^e$ Reference
\cite{MMD}
\end{table*}
\begin{figure}[b]
\resizebox{.5\textwidth}{!}{\includegraphics{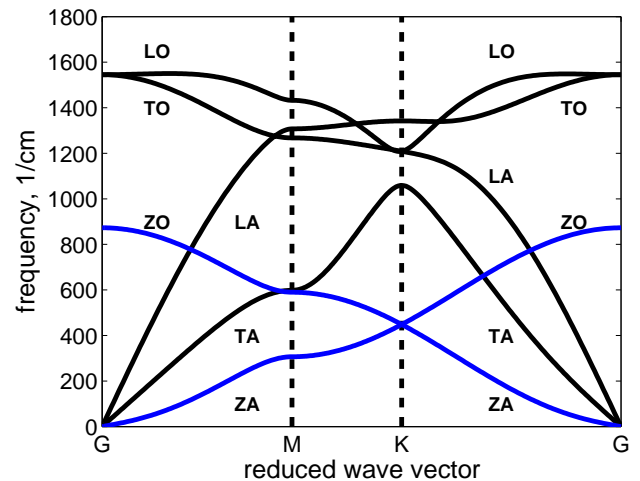}} \caption{
\label{2} Calculated phonon dispersion for graphene; the force
constants, elastic constants, and  phonon frequencies in critical
points are listed   in Tables 1, 2, and 3 correspondingly. }
\end{figure}

The calculated phonon dispersion  is shown in Fig. 2. Notice,
first, that the sound velocities (for the long waves,
$q\rightarrow\Gamma$) are isotropic in the $xy$ plane as it should
be appropriate for the  symmetry of graphene. Second, the in-plane
LO/TO modes at $\Gamma$, the in-plane LO/LA modes at $K$, and the
out-of-plane ZA/ZO modes at $K$ are doubly degenerate, because
graphene is the non-polar crystal and the $C_{3v}$ symmetry of
these points in the Brillouin zone admits the two-fold
representation (observation of splitting of that modes in graphene
would  display the symmetry braking of the crystal).

\section{Fitting to  experimental data for graphite}

Because of the lack of information on graphene, we  compare the
present theory with  experiments on graphite. In fitting to the
experimental data, we must keep in mind that  the  frequencies in
graphene for the out-of-plane branches should  be less than their
values in graphite, since the atoms are more free to move in the
$z$ direction in graphene comparatively with graphite. It is
evident that  the adjacent layers in graphite affect more
intensively the low frequencies. The interaction of the adjacent
layers can be estimated   from the observed splitting at around
130 cm$^{-1}$ [see, for instance, Ref. \cite{MM}] of the ZA and
ZO' modes  which slick together if the interaction is absent.

Thus, we have only two force constants $\alpha_z$ and $\gamma_z$
to fit four Raman frequencies of the out-of-plane modes. We see
also from Eq. (\ref{sz})  that these force constants determine the
velocity $s_z$ of the acoustic out-of-plane mode along with the
elastic constant $C_{44}$. This our result contradicts to the
well-known statement \cite{DS} that the acoustic out-of-plane mode
has a quadratic dispersion.  That can be found only if the
 condition $\gamma_z=-\alpha_z/6$ of the velocity  vanishing is
 satisfied. Therefore,  the force constant $\alpha_z$ given
 in Table 1 is obtained using  the experimental value of the out-of-plane
 Raman frequency. The small force constant $\gamma_z$ is found
 from the value of $C_{44}$, Table 2, taking the experimental error into account.
 As one can see in Table 3, the
 lower frequencies of the out-of-plane mode at other critical points
in graphene  turn out to be less than their values in graphite.
The interlayer interaction should diminish that differences. On
the other hand, the fact that the sound velocity $s_z$ is very
sensitive to the small variation of $\gamma_z$ and becomes complex
for $\gamma_z> 0.21\times 10^5$ cm$^{-2}$ indicates that graphene
is nearly unstable with respect to transformation into a phase of
the lower symmetry group at $\Gamma$.

For the in-plane modes, we have to fit eight Raman frequencies and
two elastic constants using four force constants.  The equations
(\ref{pfx}), (\ref{s12}), and (\ref{lv}) can be used as a starting
point.
 Fitting of the in-plane branches is insensitive to the imaginary
part of the constant  $\delta$. Therefore, it is taken as a real
parameter. Results of fitting are presented in Fig. 2 and in
Tables. Notice, that the extent of agreement of the present theory
with the  data obtained  for graphite corresponds to the
comparison level between the first-principal calculations  for
graphite  in Ref. \cite{MRTR} and their experimental data (see
Table 3). We see only qualitative discrepancy in the subsequence
of the levels at M:
 in Fig. 2 the highest level is the LO mode, whereas in Ref.
\cite{MRTR} the crossover  of the TO and LO modes is found out on
the $\Gamma$-M line (similar to $\Gamma$-K line), yielding the TO
mode higher at M. We examined  versions with the crossover. The
agreement with experiments is not so good in these cases as for
one that shown in Fig. 2 and Tables, but the discrepancies of the
order of 50 cm$^{-1}$ between the different experiments  as well
as the distinctions between graphene and graphite do not allow us
to choose  the version conclusively. The experiment on graphene
would clarify this point.
\section{Summary} We calculate the phonon dispersion in
graphene using the Born-von-Karman model with only the first- and
second-neighbor interactions imposed by the symmetry constraints.
 The bending (out-of-plane)
modes are not coupled with the in-plane branches and indicate the
latent instability of graphene with respect to transformation into
a lower-symmetry phase. The Raman frequencies of these modes are
less than the corresponding values in graphite. The fitting of the
higher in-plane  modes shows the good agreement of the calculated
optical frequencies as well as elastic constants with experiments.

\acknowledgements
 The work was supported by the Russian Foundation
for Basic Research  (grant No.07-02-00571).

\end{document}